# Plasmonic nanoantennas as integrated coherent perfect absorbers on SOI waveguides for modulators and all-optical switches

Roman Bruck,[1,*] and Otto L. Muskens[1]

[1] *Physics and Astronomy, Faculty of Physical Sciences and Engineering, University of Southampton, Southampton SO17 1BJ, UK*
[*]*r.bruck@soton.ac.uk*

**Abstract:** The performance of plasmonic nanoantenna structures on top of SOI wire waveguides as coherent perfect absorbers for modulators and all-optical switches is explored. The absorption, scattering, reflection and transmission spectra of gold and aluminum nanoantenna-loaded waveguides were calculated by means of 3D finite-difference time-domain simulations for single waves propagating along the waveguide, as well as for standing wave scenarios composed from two counterpropagating waves. The investigated configurations showed losses of roughly 1% and extinction ratios greater than 25 dB for modulator and switching applications, as well as plasmon effects such as strong field enhancement and localization in the nanoantenna region. The proposed plasmonic coherent perfect absorbers can be utilized for ultracompact all-optical switches in coherent networks as well as modulators and can find applications in sensing or in increasing nonlinear effects.

**OCIS codes:** (140.4780) Optical resonators; (260.3160) Interference; (310.3915) Metallic, opaque, and absorbing coatings; (160.3918) Metamaterials; (050.6624) Subwavelength structures; (230.7370) Waveguides; (130.3120) Integrated optics devices; (250.5403) Plasmonics; (290.0290) Scattering;

**1. Introduction**

Absorption in integrated optics is generally seen as a parasitic effect, which needs to be minimized. However, absorption and in particular the control of absorption is also an opportunity to actively influence the propagation of light in waveguides. An example of this is critical coupling, where light can be coupled with 100% efficiency from a waveguide into a ring resonator by matching losses [1]. Other examples where absorption is exploited are electro-absorption modulators [2]. Recently, the use of phase-change chalcogenides was demonstrated to control light absorption in a waveguide using a high-intensity laser to induce the phase-change [3].

Related to the phenomenon of critical coupling, coherent perfect absorption (CPA) was recently predicted [4,5] and demonstrated experimentally by Wan et al. [6]. By interfering two light beams in a slightly absorbing 110 µm-thick silicon wafer, they demonstrated 100% absorption due to the exact cancellation of reflected and transmitted light beams with opposite phase and equal amplitude. The phenomenon can be interpreted as the time-reversed analogy of a laser, or anti-laser [4,5,7,8]. Similar conditions leading to CPA can be achieved using ultrathin absorbers, as was demonstrated by Zhang et al. by interfering two beams on a thin plasmonic metamaterial [9]. Depending on the position of the ultrathin absorber layer in the node or antinode of the standing-wave light field, absorption could be completely suppressed or enhanced, in principle allowing perfect absorption to be achieved.

Here, we investigate the feasibility of integrating CPA devices into integrated photonic waveguides. We propose that, in the context of integrated waveguides, coherent absorption can be achieved by loading waveguides with absorbers of sub-wavelength dimensions in the propagation direction. In order to achieve CPA, the interaction of the absorber structure with the guided light in the waveguide has to be optimized to achieve the right amount of absorption with matching interference between reflected and transmitted light. We show numerically that approximate conditions for CPA can be achieved using a relatively simple array of resonant plasmonic nanoantennas.

Plasmonic devices are well known for their ability to localize light beyond the diffraction limit [10,11,12]. Strong optical resonances can be designed by tailoring the size and shape of the nanostructures in analogy to radiowave antennas. The resulting plasmonic nanoantennas have found wide applications in a variety of fields ranging from nonlinear optics to molecular biosensing [10,11,13,14]. While most applications benefit from minimizing nonradiative plasmonic losses, in recent years the strategic use of plasmonic nanostructures with designed absorption has attracted interest for applications in energy harvesting [15,16], biomedicine [17], photocatalysis [18], and nonradiative control of quantum emitters [19].

We specifically investigate arrays of plasmonic nanoantennas on top of silicon on insulator (SOI) waveguides as CPA structures. Silicon photonics is chosen for its promise as a technology for interfacing nanoelectronics and photonics on a single chip [2]. Several groups have addressed the interaction of plasmonic nanostructures with waveguides [20-28]. These works have addressed topics such as plasmon-induced transparency, Fano resonances, strong plasmon-photon interactions, and directional beaming of radiation using plasmonic Yagi-Uda arrays. So far, plasmon-induced CPA using antennas on waveguides has not yet been investigated. Compared to three-dimensional plasmonic CPA [6], a waveguide configuration offers a potentially simpler system as fewer modes will need to be matched. In particular, when considering scattering out of the waveguide as an additional (coherent) loss channel, only the coupling with the reflected and transmitted waveguide modes has to be optimized. While coherent scattering losses strictly speaking are not included in CPA, they improve the performance of our waveguide devices. As we will show, the balance between coherent absorption and scattering losses can be independently tuned to some extent by the scatterer configuration.

Our numerical studies are aimed at assessing the performance of a simple model system, consisting of gold or aluminum nanoantennas on an SOI waveguide. We address key questions such as: can sufficient absorption be achieved in an antenna on waveguide configuration? Are realistic structures with experimentally achievable design parameters capable of achieving CPA devices with a technologically interesting modulation contrast? What are the design tolerances and how broadband is the CPA effect?

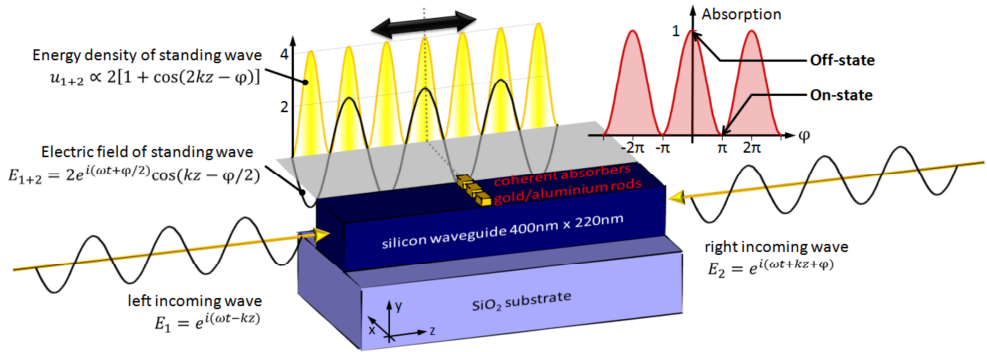

Fig. 1. Concept of coherence perfect absorption by plasmonic nanoantennas on top of silicon waveguides. Depending on the position of the antennas relative to the nodes of a standing wave in the waveguide, composed by two counterpropagating waves, absorption can be suppressed or maximized. The position of the nodes of the standing wave can be adjusted by manipulating the phase relation of the two single waves. The SiO$_2$ cladding was removed in this picture for clarity.

## 2. Method

In a scenario (see Fig. 1), where two counterpropagating waves of same wavelength and amplitude form up a standing wave in the waveguide with spatially constant nodes of minimum and maximum average energy density, the absorbance strongly depends on the position of the antennas relative to the nodes of the standing wave. Positioned in a minimum node of the standing wave, almost no absorption will take place, while positioned in a maximum node of the standing wave, absorption will be large. By manipulating the phase of at least one of the single waves, the position of the nodes can be shifted and the amount of absorption can be adjusted. For CPA, the absorber should absorb 50% of the incoming light for a single wave [9]. Since the average energy density in a maximum node of the standing wave is four times the average energy density of one of the single waves, the absorber will then absorb twice the power of a single wave, which corresponds to the total incoming power. Coherent reflection losses at the structure are cancelled for each input channel by destructive interference with the transmitted light of the other channel.

The characteristics of gold or aluminium nanoantenna array-loaded SOI waveguides were investigated by means of 3D finite-difference time-domain (FDTD) simulations [29] for single waves propagating along the waveguide, as well as for standing wave scenarios composed from two counterpropagating waves.

Simulations were performed using a non-uniform mesh with a resolution of 2 nm in the region of the antennas. Light was launched into the fundamental mode of the 400 nm wide and 220 nm thick SOI waveguide in the form of a transversal electric (TE) polarized Gaussian pulse of 1.55 µm central wavelength and 200 nm spectral width. The distribution of the injected energy was evaluated in the possible output channels i) transmission, ii) reflection, iii) out-of-plane scattering, and iv) absorption. Three different scenarios were simulated for various absorber geometries: i) single wave spectra of the absorbers, where a Gaussian pulse was launched only from one side into the waveguide, ii) spectra in the off-state (maximum absorption) of a switch/modulator where two, at the position of the antennas constructively interfering Gaussian pluses were launched from the opposite ends of the waveguide and iii) spectra in the on-state of a switch/modulator (minimum absorption) where two, at the position of the antennas destructively interfering Gaussian pluses were launched from the opposite ends of the waveguide. Destructive interference was achieved by shifting the start position of one of the Gaussian pulses by λ/2. The SOI waveguide is surrounded by SiO$_2$, which was modeled with a constant refractive index of 1.46. The material properties of the silicon waveguides and the metal antennas are based on the values given by Palik [30].

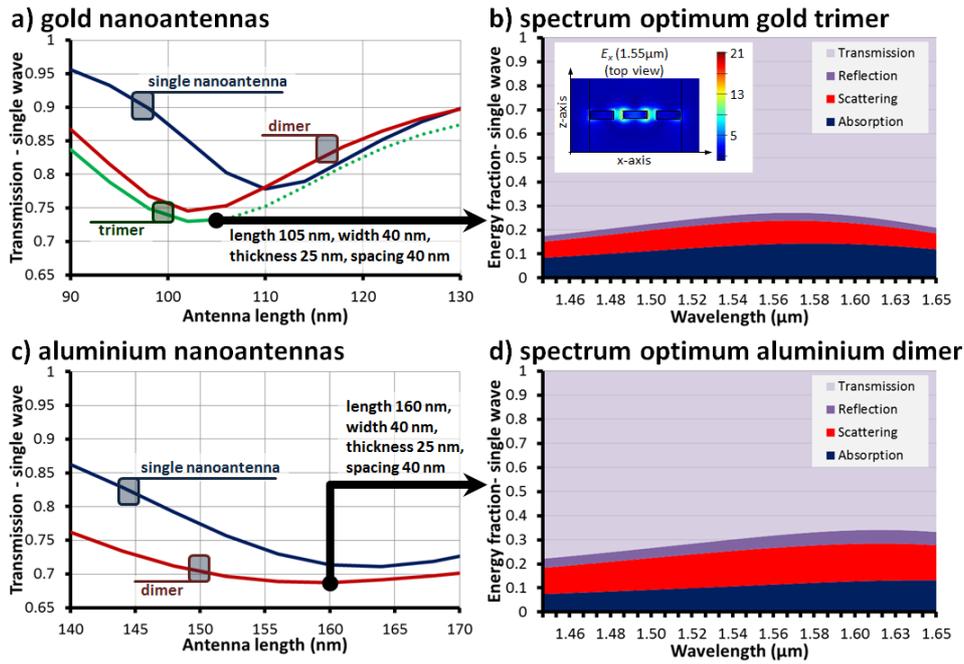

Fig. 2. Optimization of length and number of antennas for a) gold and c) aluminium antennas by minimizing transmission for a single wave ($\lambda$ = 1.55 µm, TE-polarization). The spectra of the optimum configurations are given in b) for the gold trimer and in d) for the aluminium dimer. The inset in b) shows the exited plamon resonance mode for the gold trimer.

## 3. Results

Figure 2a) and c) show the optimization of the antenna length and the number of antennas in a single row by minimizing transmission for a single wave at the design wavelength of 1.55 µm for gold and aluminium antennas, respectively. The width of the antennas of 40 nm and the thickness of the metallic layer of 25 nm were held constant. The antenna(s) were centered on the waveguide. If more than one antenna is used, the antennas were spaced by 40 nm. By employing more than one antenna, the interaction with the guided mode is increased, as more of the waveguide width is covered. The shift of the optimum resonance length from 110 nm for a single gold antenna to 105 nm for a gold trimer is a result of capacitive coupling between nanoantennas. For gold antennas, the optimum configuration was found to be three antennas (trimer) of 105 nm length, with a transmission below 75% for a single wave. Fig 2b) gives the corresponding spectra for this optimized gold structure and the inset shows the electric field ($E_x$ component) profile of the plasmon resonance. Typically, the fields within the antennas decay in less than 20 fs, thus not posing any limit to the switching speed of the device. The optimum length of aluminum antennas occurs at a larger length than their gold counterparts. The optimum configuration in aluminum was found to be a dimer composed of two antennas of 160 nm length. The performance of the aluminum antennas is otherwise comparable with gold antennas (Fig. 2c and d).

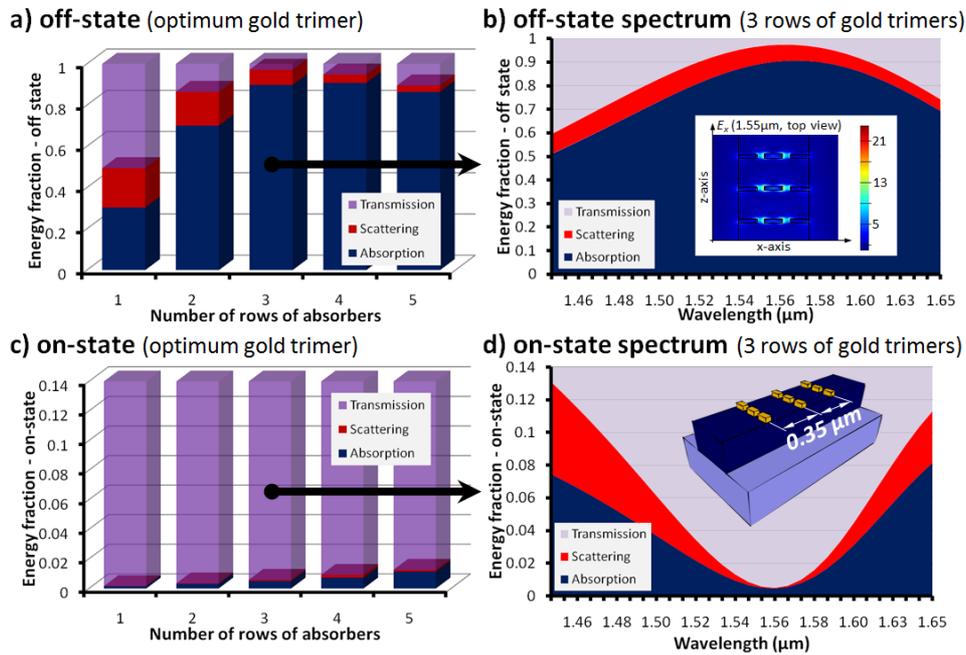

Fig. 3. Transmission, absorption and scattering for different numbers of rows of gold trimers if the rows are in a) maximum nodes (off-state) and c) in minimum nodes (on-state) of the standing wave ($\lambda = 1.55$ µm, TE). Spectra of the best configuration, consisting of three trimer rows (spaced by 0.35 µm) for the off- and the on-state are given in b) and d), respectively. The inset in b) shows the electric field profile of the plasmon resonance in the off-state, while the inset in d) depicts the simulated structure. In the off-state, the transmission drops below 3% (< -15 dB) for the design wavelength. In the on-state, losses are below 1%.

To increase absorption, additional rows of antennas were introduced into the design. Here we consider the standing wave configuration as was illustrated in Fig. 1. By choosing the row spacing corresponding to the node spacing of the standing wave at the design wavelength, all rows are either in maximum (off-state) or in minimum (on-state) nodes of the standing wave. Figure 3 compares the performance of gold absorbers for different numbers of trimer rows in the off-state (a) and the on-state (c) for $\lambda = 1.55$ µm. The row spacing for both graphs is 0.35 µm. By adding a second and a third trimer row, absorption in the off-state increases. For the optimum configuration of three rows of trimers, the spectra in the off- and the on-state are plotted in Fig. 3b) and d), respectively. At $\lambda = 1.55$ µm, transmission drops below 3% or -15 dB in the off-state, while transmission in the on-state exceeds 99%. The bandwidth of the device covers the telecommunications C-band (1.53 µm - 1.565 µm) with less than 4% transmission in the off-state and still better than 99% in the on-state.

Because of the geometry involving counterpropagating beams in a single optical mode, it is not possible to separate coherent reflection from transmission in the standing wave configuration. Therefore, reflection from the structure adds to the overall transmission of the coherent absorber device. However, the transmission in the off-state of the three-row structure is substantially lower than the reflection from a single row of trimers in Fig. 2b), indicating that coherent reflection losses are suppressed in standing wave configurations by interference of the different - reflected and transmitted – contributions in the composite device. This important property shows the CPA characteristics of the multiple row device, where the overall performance depends critically on the mutual interferences between the coupled antennas [27]. Remarkably, the performance in Fig. 3a) is reduced by adding a fourth and a fifth absorber row to the device.

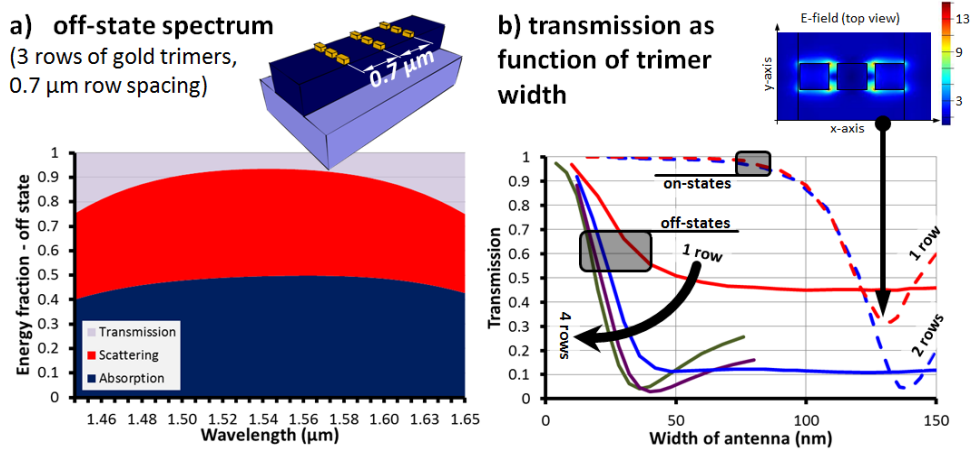

Fig. 4. a) Off-state spectra with doubled row spacing compared to Fig. 3b). b) on- and off-state transmission for different numbers of gold trimer rows as function of the width of the nanoantennas. The inset shows the electric field profile of the higher order plasmon mode, which is responsible for the dip in the on-state transmissions for 130 nm wide antennas. The on-state transmission for three and four rows are not shown, as they qualitatively identical with the other on-state curves for the range of interest (width < 80nm).

The increased transmission in the off-state can be attributed to the fact that the intensity decrease of the two counterpropagating waves over the multiple absorber structure results in a deviation from a perfect standing wave in the absorber region, which subsequently degrades the device performance. It is noteworthy that the optimum row spacing for minimum transmission in the off-state and maximum transmission in the on-state are not identical. Due do the strong coupling of the guided light with the plasmon absorbers in the off-state, the optical length of the waveguide slightly changes, thus influencing the optimum row spacing. In the simulations it was found that the optimum row spacing for minimum transmission in the off-state is 350 nm, while the optimum row spacing for maximum transmission in the on-state is around 390 nm. Since simulation results (not shown here) revealed that deviations from the optimum row spacing in the off-state have a much stronger impact on the performance than deviations from the optimum in the on-state, the row spacing was set to 350 nm.

For the results in Fig. 3, the absorber rows were placed in adjacent nodes in the standing wave, approximately 350 nm spaced. However, as long as all rows are in minimum or maximum nodes, even if not in adjacent ones, similar device behavior can be expected. In Fig. 4a) the off-state spectra for a device with 700 nm row spacing are given as comparison to Fig 3b). Interestingly, the amount of scattering is considerably larger for the 700 nm spacing. By increasing the row spacing, the absorber rows become a diffractive grating with pronounced out-of-plane diffraction, which contributes in the numerical model to the scattering channel. Thus, by tuning the spacing of the rows, it is possible to move from an absorption-dominated device to a strongly diffracting device.

To achieve CPA, the absorption "strength" of the structure needs to be balanced. If absorption strength becomes too strong, as was seen earlier by adding a fourth and fifth absorber row, the performance of the device starts to decrease. In addition to the number of rows, the absorption strength of each individual row can be tailored by adjusting the volume of the antennas in which they interact with the standing wave. This gives an additional degree of freedom for the design of such a device, which is analyzed in Fig. 4 b). The interaction volume of the antennas is hereby adjusted by changing the width of the antennas. The plot shows that for a single row, off-state transmission decreases with increasing antenna width. However, regardless of the antenna width, the maximum absorption strength for a single row is limited, thus transmission reaching a plateau around 45 %. A similar picture can be found for the two row absorber. Sufficient absorption cannot be achieved with a two row absorber of any antenna width. Thus, transmission reaches a plateau again, even if much lower than for the one row absorber. This picture changes if the absorbing structure is composed of more than two rows. For small antenna widths, transmission decreases for multi-row absorbers if the width is increased. Then an optimum antenna width is reached, which depends on the number of the rows (e.g. 40 nm for three rows and 36 nm for four rows), before an additional increase in width, i.e. absorber strength, leads to an increase in transmission. The on-state transmissions are almost stable for antenna widths up to 60 nm. This is valid for all number of rows. Then, the antennas become too wide and start to penetrate outside the regions of low energy density of the standing wave. For 130 nm wide antennas, on-state transmission is further suppressed by coupling to a higher-order mode of the structure, illustrated by the near-field map in the inset.

Other applications for plasmonic antennas on waveguides include the increase of non-linear effects or other effects that can be increased by the field enhancement of the antennas. Simulations revealed that the effective area $A_{eff}$ for cladding non-linearity, as defined in reference [31], is two orders of magnitude smaller in the region where waveguide is loaded with gold trimers. The electric field between the antennas is strongly enhanced and more than an order of magnitude larger than in the center of the single-mode waveguide without antennas (cf. Fig. 2b). This enhancement may be of interest for applications such as optical sensing using the waveguide to efficiently interface light with the plasmonic nanostructure. The combination of CPA and near field enhancement for optical sensing lies beyond the scope of current work and will be addressed in future studies.

## 3. Modulator and all-optical switch concept

In principle, the device presented in Fig. 1 can be used as an all-optical switch with an extremely compact footprint. The length of the switch would be the length of the absorber structure (< 1μm for an absorber with three rows). However, this design bears the difficulty that the outgoing light travels in the same waveguide as the incoming light. For separation of incoming and outgoing light into different waveguides, one would have to rely on optical circulators [32], which would unnecessarily complicate the device design.

Therefore, we propose to integrate the CPA structure in the middle of an evanescent X-coupler, as depicted in Fig. 5a). Evanescent couplers are well known elements in integrated optics, where the overlap of the evanescent tail of a guided mode in one waveguide with an adjacent waveguide induces a power transfer between the waveguides. If the length of the coupling region is chosen correctly, typically a few ten micrometers for SOI waveguides, a 100% power transfer can be achieved, generating a so called X-coupler, where injected light always leaves at the diagonal port. If two counterpropagating waves are inserted in into the X-coupler, they will form a standing wave in the middle section of both waveguides of the X-coupler, where 50% of each individual wave is transferred to the adjacent waveguide. By placing identical absorbers on both waveguides of the X-coupler, this device will behave identically as the device given in Fig. 1, with the difference that the transmitted light will be output in a different waveguide. Since the coupling length of the coupler is much larger than the width of the absorber, the positioning of the absorber along the coupler is not of vital importance.

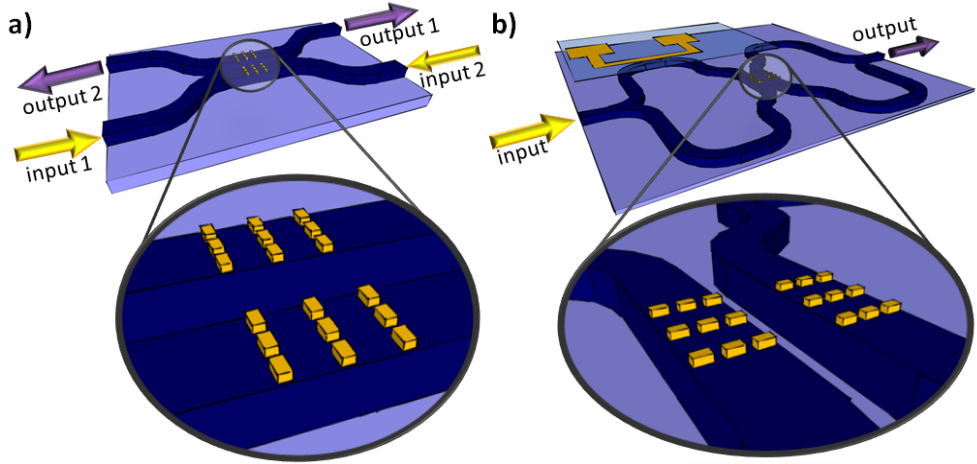

Fig. 5. Proposed device concepts for a) an all-optical switch and b) a modulator employing antennas on top of the waveguides as coherent absorbers in the middle section of evanescent X-coupler, where the condition of two counterpropagating waves of same amplitude is fulfilled locally. In b) an electro-optic or thermo-optic actuator for the modulator is indicated.

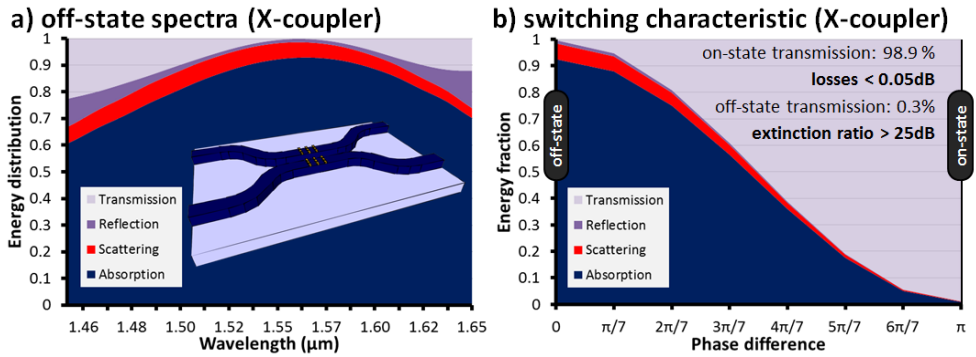

Fig. 6. a) off-state spectra of X-coupler with gold absorbers (three trimer rows) on each waveguide for $\lambda = 1.55$ µm. The inset shows the simulated structure. b) The switching characteristic of the X-coupler switch perfectly follows the expected sinusoidal curve. The performance figures are indicated in the graph.

Simulation results for an X-coupler with three rows of gold trimer antennas are presented in Fig. 6. The two 11.6 µm long parallel waveguides of the X-coupler are spaced by a 200 nm gap. The off-state spectra of this device is shown in Fig. 6a), while Fig. 6b) gives the switching characteristic, which perfectly follows the expected sinusoidal characteristic. The complete device shows very small losses in the on-state of roughly 1% or less than 0.05 dB. The extinction ratio, i.e. the ratio between on-state and off-state transmission, was calculated to be larger than 25 dB.

This concept can be further developed into a modulator structure, by feeding both inputs of the all-optical switch from the same source, as proposed in Fig. 5b). By shifting the phase of one of the two inputs by π, e.g. by a thermo-optic or electro-optic element, the modulator can be switched into the on- or the off-state.

Despite certain similarities with the Mach-Zehnder interferometer (MZI) modulator, the proposed modulator has different output characteristics. Essentially, in an MZI modulator, the propagating light can only be routed from one output to the other. In the proposed CPA modulator both outputs are switched on or off simultaneously. This is fundamentally different from MZI modulators, as the light in the off-state is not simply rerouted but converted into heat, thus taking it out of the optical circuit and preventing it from further propagation in the device as unwanted, unguided stray light, which can degrade device performance.

**4. Conclusion**

We proposed to utilize of losses in resonant plasmonic structures to induce coherent perfect absorption (CPA) in waveguide structures. CPA can be employed for integrated, waveguide-based modulators and all-optical switches. Concepts for these devices were given, where the absorbers are integrated in an evanescent X-coupler. 3D FDTD simulations of these devices loaded with three rows of gold trimer antennas on top of silicon waveguides revealed off-state transmissions of only 0.3%, while maintaining almost lossless transmission (~99%) in the on-state ($\lambda = 1.55$ µm, TE-polarization). This corresponds to an extinction ratio of more than 25 dB. The calculated absorbers have a length of only 740 nm and show a flat and broadband spectral characteristic throughout telecommunication bands. As the absorption in the antennas takes place in less than 20 fs, no restriction to the switching speed of the devices is expected.

The proposed modulator and all-optical switch can find applications in telecommunication applications and in coherent networks as well as in sensing or in increasing nonlinear effects. The large absorption of the multi row antenna device means that light can be efficiently removed from the optical circuit. In a static configuration, the CPA device concept can thus be exploited as an efficient beam dump, or in combination with a Schottky diode, as an on-chip plasmonic photodetector [33]. While the current studies did not specifically consider the device stability under photoabsorption, an area of potential interest is the possibility to achieve nonlinear and/or bistable devices [34] exploiting the combination of CPA, plasmonic local field enhancement and highly localized photoabsorption around the metal nanostructures. Such an intrinsically nonlinear CPA device might ultimately remove the need for an external phase shifter in the above designs, opening up new avenues for ultracompact integrated photonic circuits.

**Acknowledgements**
The authors acknowledge support from EPSRC through grant no. EP/J016918.